\documentclass[]{interact}

\usepackage{placeins}
\usepackage{longtable}
\usepackage{multirow}
\usepackage{soul,color}
\usepackage{epstopdf}
\usepackage[caption=false]{subfig}
\usepackage[doublespacing]{setspace}

\sethlcolor{yellow} 

\usepackage[numbers,sort&compress]{natbib}
\bibpunct[, ]{[}{]}{,}{n}{,}{,}
\renewcommand\bibfont{\fontsize{10}{12}\selectfont}
\makeatletter
\def\NAT@def@citea{\def\@citea{\NAT@separator}}
\makeatother

\theoremstyle{plain}

\theoremstyle{definition}

\theoremstyle{remark}

\begin{document}

\title{MolMod: An open access database of force fields for molecular simulations of fluids}

\author{
\name{Simon Stephan,\textsuperscript{a}\thanks{E-mail: simon.stephan@mv.uni-kl.de} Martin Thomas Horsch,\textsuperscript{b} Jadran Vrabec,\textsuperscript{c} Hans Hasse\textsuperscript{a}}
\affil{\textsuperscript{a}Technische Universit\"at Kaiserslautern, Laboratory of Engineering Thermodynamics (LTD), Erwin-Schr\"odinger-Str.\ 44, 67663 Kaiserslautern, Germany\\
\textsuperscript{b}UK Research and Innovation, STFC Daresbury Laboratory, Keckwick Lane, Daresbury, Cheshire WA4 4AD, United Kingdom\\
\textsuperscript{c}Technische Universit\"at Berlin, Thermodynamics and Process Engineering, Stra\ss{}e des 17.\ Juni 135, 10623 Berlin, Germany}
}

 \maketitle

\begin{abstract}
	The MolMod database is presented, which is openly accessible at http://molmod.boltzmann-zuse.de/ and contains presently intermolecular force fields for over 150 pure fluids. It was developed and is maintained by the Boltzmann-Zuse Society for Computational Molecular Engineering (BZS). The set of molecular models in the MolMod database provides a coherent framework for molecular simulations of fluids. The molecular models in the MolMod database consist of Lennard-Jones interaction sites, point charges, and point dipoles and quadrupoles, which can be equivalently represented by multiple point charges. The force fields can be exported as input files for the simulation programs \textit{ms}2 and \textit{ls}1 mardyn, Gromacs, and LAMMPS. To characterise the semantics associated with the numerical database content, a force-field nomenclature is introduced that can also be used in other contexts in materials modelling at the atomistic and mesoscopic levels. The models of the pure substances that are included in the data base were generally optimised such as to yield good representations of experimental data of the vapour-liquid equilibrium with a focus on the vapour pressure and the saturated liquid density. In many cases, the models also yield good predictions of caloric, transport, and interfacial properties of the pure fluids. For all models, references to the original works in which they were developed are provided. The models can be used straightforwardly for predictions of properties of fluid mixtures using established combination rules. Input errors are a major source of errors in simulations. The MolMod database contributes to reducing such errors.
\end{abstract}


\section{Introduction}

Reliable information on thermophysical properties is of fundamental importance for the development and optimisation of processes in chemical engineering, but also in many other fields, such as energy conversion and storage. With the increasing availability of computational resources, simulation techniques are today an attractive alternative to classical experiments in laboratories to obtain accurate thermophysical properties \cite{Maginn_2009,Economou_2017,Vrabec_2018}. Compared to laboratory experiments, computer experiments have the advantage, that problematic substances (toxic, flammable, explosive etc.) and extreme conditions can be studied without any problems. Computer experiments, like molecular dynamics (MD) and Monte Carlo (MC), have therefore become an accepted complement and sometimes even substitute to traditional experimental work. Therefore many scientific journals that were afore focussed on reporting experimental data on thermophysical properties have started recently to accept and also call for corresponding molecular simulation data \cite{Kofke_2016}. Especially phase equilibria, transport and interfacial properties of pure fluids and fluid mixtures are of special interest in chemical engineering and also in many other fields. Knowledge of these thermophysical properties is often required for a wide range of temperature and pressure and with a high accuracy. The accuracy of thermophysical property obtained by molecular simulation, in the sense that the simulation shall reproduce properties of real substances, depends mainly on the applied force field. The development of suitable force fields is therefore of prime importance for the success of molecular methods in materials modelling. Moreover, it is not sufficient that good force field are developed. They must also be made available to potential users in a way that they can be handled easily in the simulation package with which the user wants to carry out his work. This is of special importance, as input errors are known to be one of the major sources of simulation errors\cite{Schappals_2017}. The present work addresses this challenge. 

Generalised force fields, such as CHARMM  \cite{Miller_2008,Vanommeslaeghe_2010,Vanommeslaeghe_2012_A,Vanommeslaeghe_2012_B}, AMBER \cite{Cornell_1995,Wang_2004,Wang_2006}, or GROMOS \cite{Scott_1999,Oostenbrink_2004}, are mainly used for biological systems, and focus therefore on aqueous systems at near-ambient conditions. Such generalised force fields mostly do not satisfy the accuracy that is required in chemical engineering for data on phase equilibrium, transport and interfacial properties. Force fields that are particularly aiming at an accurate description of phase equilibria are for example OPLS \cite{Jorgensen_2005,Jorgensen_1996,Dodda_2017}, TraPPE \cite{Eggimann_2014}, and OPPE \cite{Ungerer_2000,Bourasseau_2003,Delhommelle_2000}. They are transferable force fields which means that the interaction parameters of chemical groups or atoms can be transferred among substance families. 

The molecular models of the pure substances in the MolMod database \cite{MolMod_website} were tailored to describe thermophysical properties of that substance. Only fluid properties are considered. The interaction parameters of a molecular species are only valid for that very substance, i.e. no extrapolation to other substances is intended. The advantage of this individual parametrisation lies in the high accuracy of each model which can be achieved. Furthermore, it is known that using such models, extrapolations to other thermophysical properties and other conditions than those that were used in the parametrisation are often successful \cite{Huang_2011,Merker_2010,Schnabel_2007c,Eckl_2008a}. The models can also be applied for studying fluid mixtures using well-established combination rules \cite{Schnabel_2007a}. The force fields in the MolMod database were developed under the auspices of the Boltzmann-Zuse Society of Computational Molecular Engineering (BZS) as described in more than 30 original peer-reviewed publications. The molecular models were parametrised in a consistent way with respect to vapour-liquid equilibrium data, which is therefore described very well. Other pure component properties, such as homogeneous $pvT$ data, caloric properties, speed of sound, interfacial properties, predicted from these models were found to be in good agreement with experimental data -- where available \cite{Eckl_2008,Schnabel_2007c,Guevara-Carrion_2016,Thol_2015,Werth_2017_B,Becker_2016,Stephan_2018_D}. The MolMod database currently consists of molecular models of approximately 150 pure substances of mainly moderately sized molecules that are modelled as rigid bodies. Also some models for ions are included in the database. The molecular models in the MolMod database consist of Lennard-Jones interaction sites, point charges, and point dipoles and quadrupoles. 

The models can also be combined to model mixtures. Usually the Lorentz-Berthelot combination rules are found to yield acceptable predictions of the behaviour of binary and multicomponent mixtures. However, the accuracy is then often below that which is achieved for the pure substances. To obtain similar accuracy also for mixtures, it is often sufficient to adjust a single state-independent binary parameter in the Berthelot combination rule for the dispersive cross-interaction \cite{Vrabec_2009,Schnabel_2007a}.

The core of the MolMod database is a download feature for several molecular simulation codes. The codes that are supported in the present version are: LAMMPS \cite{Plimpton_95}, Gromacs \cite{VanDerSpoel_2005}, \textit{ms}2 \cite{Rutkai_2017}, and \textit{ls}1 mardyn \cite{Niethammer_2014}. Databases that provide input files for molecular interactions and generalised force fields have already been introduced in the fields of microbiology and chemistry \cite{VanDerSpoel_2012,Hermjakob_2004,Zanzoni_2002,Irwin_2005,Wang_2006,Vanommeslaeghe_2012_A,Vanommeslaeghe_2012_B}, but to the best of our knowledge not for individually parametrised highly accurate molecular force fields, focussing on the description of thermophysical properties of fluids.

\section{The MolMod Database}

\begin{figure}
	\centering
	\includegraphics[width=0.55\linewidth]{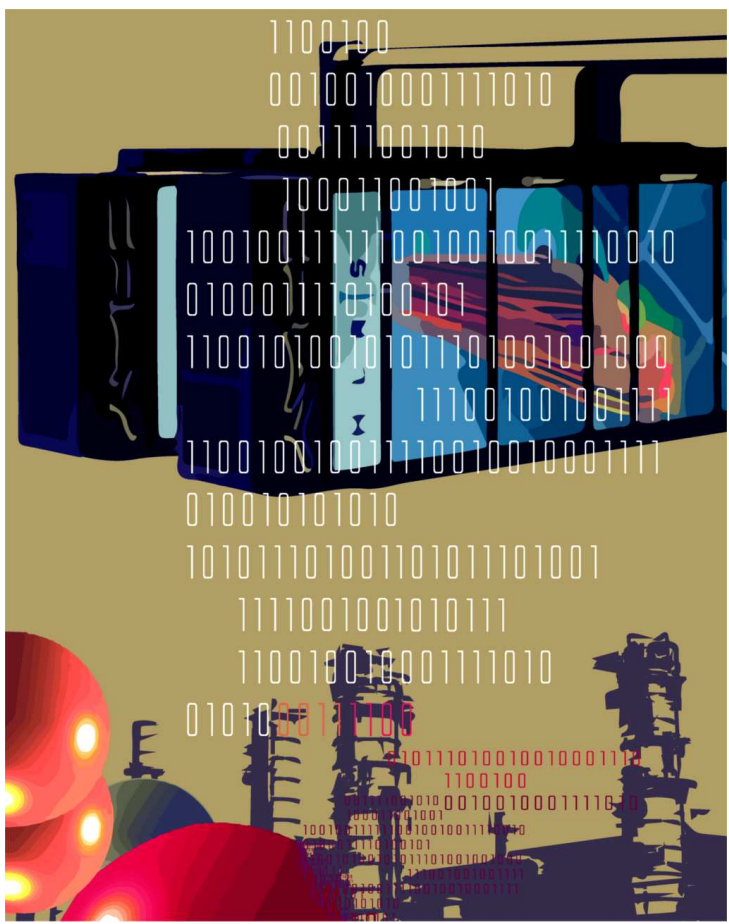}
	\caption{MolMod database web frontend artwork: http://molmod.boltzmann-zuse.de/.}
	\label{fig:Pic3}
\end{figure}

The MolMod database http://molmod.boltzmann-zuse.de (Web frontend cf. Fig. \ref{fig:Pic3}) provides a range of search functionalities, e.g. for substances, CAS-numbers, model classes etc. Each molecular model has an individual page, which contains the full specification of the molecular model and the download links for the input files. Also, each molecular model is hyperlinked to the original publication. The database contains a comprehensive nomenclature section that describes the specifications in the input files for all supported simulation programs to facilitate the simulators integrating the force fields in their simulation workflows. Known misprints in publications are clearly emphasised in the database. 

All force fields in the MolMod database are currently rigid single- or multi-site models, consisting of Lennard-Jones 12-6 interaction sites, point charges, point dipoles, and point quadrupoles. All supported interaction potentials are pairwise additive. 

In all models, the repulsion and dispersion between two particles $i$ and $j$ at a distance $r_{ij}$ is described by the standard Lennard-Jones 12-6 potential:
\begin{equation}
u_{ij}^\mathrm{LJ}(r)=4\varepsilon \Big[
\left(\frac{\sigma}{r_{ij}}\right)^{12}-\left(\frac{\sigma}{r_{ij}}\right)^6 \Big]\, ,
\end{equation}
where $\varepsilon$ is the dispersion energy and $\sigma$ the size parameter of an interaction site. The interaction between two different Lennard-Jones sites are described by the Lorentz-Berthelot combination rules \cite{Lorentz_1881,Berthelot_1898}. 

The electrostatic interactions in the database are modelled by point charges, point dipoles, and point quadrupoles. The latter two describe the electrostatic field of two and four point charges of the same magnitude, respectively. The advantage of higher order polarities in molecular simulations lies in a speed-up of their computation of up to 60\% compared to the a corresponding arrangement of point charges \cite{Deublein_2011}. It has also been claimed that multipoles give a better description of electrostatic interactions \cite{Stone_2008}. 

The electrostatic potential between two point charges $q_i$ and $q_j$ is given by Coulomb's law
\begin{equation}
u^\mathrm{ee}_{ij}(r_{ij})=\frac{1}{4\epsilon_0\pi}\frac{q_iq_j}{r_{ij}}\, ,
\end{equation}
where $q$ is the magnitude of the charge and $r_{ij}$ the distance between two point charges.

A point dipole describes the electrostatic field of two point charges with equal magnitude, but opposite sign at a mutual distance $a\rightarrow 0$. Its dipole moment magnitude is defined by $\mu=qa$. The electrostatic potential between two point dipoles with the moments $\mu_i$ and $\mu_j$ at a distance $r_{ij}$ is given by \cite{Garzon_1994,Gray_1984_book,Stoll_2001}
\begin{equation}
u_{ij}^\mathrm{dd}(r_{ij},\theta_i,\theta_j,\phi_{ij},\mu_i,\mu_j)=\frac{1}{4\pi\epsilon_0}\frac{\mu_i\mu_j}{r^3_{ij}}\Big[(\sin(\theta_i)\sin(\theta_j)\cos(\phi_{ij})-2\cos(\theta_i)\cos(\theta_j)\Big],
\end{equation}
where the angles $\theta_i$, $\theta_j$ and $\phi_{ij}$ indicate the relative angular orientation of the two point dipoles 
with $\theta$ being the angle between the dipole direction and the distance vector of the two interacting dipoles and $\phi_{ij}$  being the azimuthal angle of the two dipole directions.

A linear point quadrupole describes the electrostatic field induced either by two collinear point dipoles with the same moment, but opposite orientation at a distance $a\rightarrow 0$ or three point charges, e.g. ($q$, $-2q$, $q$). The quadrupole magnitude $Q$ is defined as $Q=2qd^2$, where $q$ is the characteristic magnitude of the three charges, and $d$ is the distance between each of the outer charges (magnitude $q$) and the $-2q$ charges in the centre. The interaction potential is given by \cite{Garzon_1994,Gray_1984_book,Stoll_2001}
\begin{align}
u_{ij}^\mathrm{qq}(r_{ij},\theta_i,\theta_j,\phi_{ij},Q_i,Q_j) = \frac{1}{4\pi\epsilon_0}\frac{3}{4}\frac{Q_iQ_j}{r^5_{ij}} \Big[ 1-5 (\cos(\theta_i)^2+\cos(\theta_i)^2 ) \\ 
-15 ( \cos(\theta_i)) ^2(\cos(\theta_j))^2+2(\sin(\theta_i)\sin(\theta_j)\cos(\phi_{ij})-4\cos(\theta_i)\cos(\theta_j))^2  \Big] \nonumber \, ,
\end{align}
where the angles $\theta_i$, $\theta_j$ and $\phi_{ij}$ indicate the relative angular orientation of the two point quadrupoles. The interaction potentials between unlike electrostatic site types can also be derived from the laws of electrostatics straightforwardly. For a detailed derivation of these equations, cf. \textit{Gray and Gubbins} \cite{Gray_1984_book}

For mixtures of fluids from the MolMod database, the modified Lorentz-Berthelot combination rules proved to give an accurate description of fluid mixtures\cite{Vrabec_2009,Schnabel_2007a,Merker_2012b}:
\begin{eqnarray}
\sigma_{ij} =\eta \frac{\sigma_i + \sigma_j}{2},\\
\varepsilon_{ij} = \xi \sqrt{\varepsilon_i\, \varepsilon_j},
\end{eqnarray}
where $\xi$ and $\eta$ are state-independent adjustable binary interaction parameters to explicitly consider the non-ideality of the respective mixture. Binary interaction parameters $\xi$ adjusted to binary vapour-liquid equilibria data \cite{Vrabec_2009,Schnabel_2005} are available for approximately 400 mixtures of molecular models from the MolMod database. It could be shown that the unmodified Lorentz combination rule, i.e. $\eta = 1$ is a good approximation in most cases \cite{Schnabel_2007a}. Using the input files provided by the MolMod database for the simulation of mixtures with the programs \textit{ms}2 and \textit{ls}1 mardyn is straightforward\cite{Rutkai_2017,Niethammer_2014} as their syntax only requires the additional specification of the mole fractions and eventually the binary interaction parameters for mixtures. Numerical values for the binary parameters $\xi$ will be included in the database by future work.

The parametrisation of molecular models that are part of the MolMod database was usually carried out in two steps: The geometry and charge distribution of multi-site models was obtained from quantum chemical calculations. The remaining intermolecular interactions were then optimised to vapour-liquid equilibrium data, i.e. saturated liquid density, vapour pressure, and enthalpy of vaporisation. Also, automated simulation and optimisation workflows have been employed, operating in the parameter space locally \cite{Deublein_2013,Merker_2012} as well as globally \cite{Kraemer_2014}, which leads to a significant reduction of the effort.  Multicriteria optimisation based on the analysis of a Pareto set has also been employed for the parametrisation lately \cite{Stoebener_2014,Stoebener_2016,Kohns_2017} to explicitly consider conflicting objectives. The agreement with experimental data is typically found to be within 4\% for the vapour pressure, 0.5\% for the saturated liquid density, and 3\% for the enthalpy of vaporisation. Predicted thermophysical properties that were not considered during the parametrisation agree with experimental data approximately 10\% for transport properties, and 20\% for surface tension data. In some cases, the deviations of the model behaviour from the available experimental data is smaller than the experimental uncertainty for a wide variety of thermophysical properties \cite{Eckl_2008a}. Some of the molecular models have also been employed to complement experimental data sets for the development of high-accuracy equations of state \cite{Thol_2015,Thol_2016A} for engineering applications.

Table \ref{tab:list of substances} summarises the molecular substances that are currently available in the MolMod database, sorted by the number of carbon atoms. The database contains molecular models of different substance classes, e.g. organic substances, noble gases, and chlorofluorocarbons. Organic substances in the database are from different chemical families, like alkanes, alkenes, alcohols, aromatics, amines, ketones, nitriles, acids, etc. Molecular models for numerous substances of fundamental importance for the chemical industry are covered, such as methanol, ethanol, formic acid, ammonia, formaldehyde, hydrogen cyanide, acetonitrile, ethylene oxide, toluene, ethylene glycol, acetone, etc. A second focus of the MolMod database comprises important refrigerants, of which it contains approximately 30, such as R134a, R22, R12, R227ea. The MolMod database also contains molecular models of substances that are gases at normal conditions and are for example important in environmental science, such as O$_2$, N$_2$, CO, CO$_2$, N$_2$O, H$_2$, H$_2$O, SO$_2$, or Ar. Also some models for ions are available in the MolMod database \cite{Reiser_2014,Deublein_2012a,Deublein_2012}. 

\begin{longtable}{llll}
	\caption{List of molecular force fields currently available in the MolMod database.}	\label{tab:list of substances} 
	\toprule 
	Substance & CAS-No. & Name & Reference \\
	\hline
	\endfirsthead
	\toprule
	Substance & CAS-No. & Name & Reference \\
	\hline
	\endhead
	\hline 
	\multicolumn{4}{r}{\textit{Continued on next page}} \\
	\endfoot
	\endlastfoot
	Ar  & 7440-37-1 & Argon & \cite{Vrabec_2001,Vrabec_2006} \\
	Br$_2$ & 7726-95-6 & Bromine & \cite{Vrabec_2001,Stoebener_2016} \\
	Cl$_2$ & 7782-50-5 & Chlorine & \cite{Vrabec_2001} \\
	F$_2$ & 7782-41-4 & Fluorine & \cite{Vrabec_2001,Stoebener_2016} \\
	H$_2$ & 1333-74-0 & Hydrogen & \cite{Koester_2018} \\
	H$_2$O & 7732-18-5 & Water & \cite{Huang_2012} \\
	H$_4$N$_2$ & 302-01-2 & Hydrazine & \cite{Elts_2012} \\
	HCl & 7647-01-0 & Hydrochloric acid & \cite{Huang_2011} \\
	I$_2$ & 7553-56-2 & Iodine & \cite{Vrabec_2001} \\
	Kr  & 7439-90-9 & Krypton & \cite{Vrabec_2001,Vrabec_2006} \\
	N$_2$ & 7727-37-9 & Nitrogen & \cite{Vrabec_2001,Stoebener_2016} \\
	N$_2$O & 10024-97-2 & Nitrous Oxide & \cite{Kohns_2017} \\
	Ne  & 7440-01-9 & Neon & \cite{Vrabec_2001} \\
	NH$_3$ & 7664-41-7 & Ammonia & \cite{Eckl_2008} \\
	O$_2$ & 7782-44-7 & Oxygen & \cite{Vrabec_2001,Stoebener_2016} \\
	SF$_6$ & 2551-62-4 & Sulfur hexafluoride & \cite{Vrabec_2001} \\
	SO$_2$ & 7446-09-5 & Sulfur dioxide & \cite{Eckl_2008b} \\
	Xe  & 7440-63-3 & Xenon & \cite{Vrabec_2001,Vrabec_2006} \\
	CBr$_2$F$_2$ & 75-61-6 & Dibromodifluoromethane & \cite{Stoll_2003} \\
	CBrCl$_3$ & 75-62-7 & Bromotrichloromethane & \cite{Stoll_2003} \\
	CBrClF$_2$ & 353-59-3 & Bromochlorodifluoromethane & \cite{Stoll_2003} \\
	CBrF$_3$ (R13B1) & 75-63-8 & Bromotrifluoromethane & \cite{Stoll_2003} \\
	CCl$_2$O & 75-44-5 & Phosgene & \cite{Huang_2011} \\
	CCl$_4$ & 56-23-5 & Carbon tetrachloride & \cite{Vrabec_2001,Guevara-Carrion_2016} \\
	CClN & 506-77-4 & Cyanogen chloride & \cite{Miroshnichenko_2013} \\
	CF$_2$Cl$_2$ (R12) & 75-71-8 & Dichlorodifluoromethane & \cite{Stoll_2003} \\
	CF$_3$Cl (R13) & 75-72-9 & Chlorotrifluoromethane & \cite{Stoll_2003} \\
	CF$_4$ & 75-73-0 & Tetrafluoromethane & \cite{Vrabec_2001} \\
	CFCl$_3$ (R11) & 75-69-4 & Trichloromonofluoromethane & \cite{Stoll_2003} \\
	CH$_2$Br$_2$ & 74-95-3 & Dibromomethane & \cite{Stoll_2003} \\
	CH$_2$BrCl & 74-97-5 & Bromochloromethane & \cite{Stoll_2003} \\
	CH$_2$Cl$_2$ & 75-09-2 & Methylene chloride & \cite{Stoll_2003} \\
	CH$_2$F$_2$ (R32) & 75-10-5 & Difluoromethane & \cite{Stoll_2003} \\
	CH$_2$I$_2$ & 75-11-6 & Methylene iodide & \cite{Stoll_2003} \\
	CH$_2$O & 50-00-0 & Formaldehyde & \cite{Eckl_2008b} \\
	CH$_2$O$_2$ & 64-18-6 & Formic acid & \cite{Schnabel_2007b} \\
	CH$_3$Br & 74-83-9 & Methyl bromide & \cite{Stoll_2003} \\
	CH$_3$Cl & 74-87-3 & Methyl chloride & \cite{Stoll_2003} \\
	CH$_3$F (R41) & 593-53-3 & Methyl fluoride & \cite{Stoll_2003} \\
	CH$_3$I & 74-88-4 & Methyl iodide & \cite{Stoll_2003} \\
	CH$_3$NO$_2$ & 75-52-5 & Nitromethane & \cite{Eckl_2008b} \\
	CH$_4$ & 74-82-8 & Methane & \cite{Vrabec_2001,Vrabec_2006} \\
	CH$_4$O & 67-56-1 & Methanol & \cite{Schnabel_2007c} \\
	CH$_5$N & 74-89-5 & Methylamine & \cite{Schnabel_2008} \\
	CH$_6$N$_2$ & 60-34-4 & Monomethylhydrazine & \cite{Elts_2012} \\
	CHBr$_3$ & 75-25-2 & Bromoform & \cite{Stoll_2003} \\
	CHCl$_2$F & 75-43-4 & Dichlorofluoromethane & \cite{Stoll_2003} \\
	CHCl$_3$ & 67-66-3 & Chloroform & \cite{Stoll_2003} \\
	CHF$_2$Cl (R22) & 75-45-6 & Difluorochloromethane & \cite{Stoll_2003} \\
	CHF$_3$ (R23) & 75-46-7 & Fluoroform & \cite{Stoll_2003} \\
	CHN & 74-90-8 & Hydrogen cyanide & \cite{Eckl_2008b} \\
	CO  & 630-08-0 & Carbon monoxide & \cite{Stoll_2003} \\
	CO$_2$ & 124-38-9 & Carbon dioxide & \cite{Vrabec_2001,Merker_2010} \\
	CS$_2$ & 75-15-0 & Carbon disulfide & \cite{Vrabec_2001} \\
	C$_2$Br$_2$F$_4$ & 124-73-2 & 1,2-Dibromotetrafluoroethane & \cite{Stoll_2003} \\
	C$_2$BrF$_3$ & 598-73-2 & Bromotrifluoroethylene & \cite{Stoll_2003} \\
	C$_2$Cl$_2$F$_4$ (R114) & 76-14-2 & 1,2-dichloro-1,1,2,2-tetrafluoro-Ethane & \cite{Stoll_2003} \\
	C$_2$Cl$_3$F$_3$ (R113) & 76-13-1 & 1,1,2-trichloro-1,2,2-trifluoro-Ethane & \cite{Stoll_2003} \\
	C$_2$Cl$_4$ & 127-18-4 & Tetrachloroethylene & \cite{Vrabec_2001,Stoebener_2016} \\
	C$_2$Cl$_4$F$_2$ (R112a) & 76-11-9 & 1,1,1,2-Tetrachloro-2,2-difluoroethane & \cite{Stoll_2003} \\
	C$_2$ClF$_3$ & 79-38-9 & Chlorotrifluoroethene & \cite{Stoll_2003} \\
	C$_2$ClF$_5$ (R115) & 76-15-3 & Pentafluoroethyl chloride & \cite{Stoll_2003} \\
	C$_2$F$_4$ & 116-14-3 & Tetrafluoroethylene & \cite{Vrabec_2001,Stoebener_2016} \\
	C$_2$F$_6$ & 76-16-4 & Perfluoroethane & \cite{Vrabec_2001} \\
	C$_2$H$_2$ & 74-86-2 & Acetylene & \cite{Vrabec_2001,Stoebener_2016} \\
	C$_2$H$_2$Cl$_3$F & 27154-33-2 & Ethane, trichlorofluoro- & \cite{Stoll_2003} \\
	C$_2$H$_2$Cl$_4$ & 630-20-6 & 1,1,1,2-Tetrachloroethane & \cite{Stoll_2003} \\
	C$_2$H$_2$F$_2$ & 75-38-7 & 1,1-Difluoroethene & \cite{Stoll_2003} \\
	C$_2$H$_2$F$_4$ (R134) & 359-35-3 & 1,1,2,2-tetrafluoro-Ethane & \cite{Stoll_2003} \\
	C$_2$H$_2$F$_4$ (R134a) & 811-97-2 & Norflurane & \cite{Stoll_2003} \\
	C$_2$H$_3$Cl & 75-01-4 & Vinyl chloride & \cite{Stoll_2003} \\
	C$_2$H$_3$Cl$_2$F (R141b) & 1717-00-6 & 1,1-Dichloro-1-fluoroethane & \cite{Stoll_2003} \\
	C$_2$H$_3$Cl$_3$ & 71-55-6 & Methylchloroform & \cite{Stoll_2003} \\
	C$_2$H$_3$Cl$_3$ & 79-00-5 & 1,1,2-Trichloroethane & \cite{Stoll_2003} \\
	C$_2$H$_3$ClF$_2$ (R142b) & 75-68-3 & 1-chloro-1,1-difluoro-Ethane & \cite{Stoll_2003} \\
	C$_2$H$_3$F & 75-02-5 & Vinyl fluoride & \cite{Stoll_2003} \\
	C$_2$H$_3$F$_3$ (R143a) & 420-46-2 & 1,1,1-trifluoro-Ethane & \cite{Stoll_2003} \\
	C$_2$H$_3$N & 75-05-8 & Acetonitrile & \cite{Eckl_2008b,Deublein_2013} \\
	C$_2$H$_4$ & 74-85-1 & Ethylene & \cite{Vrabec_2001,Stoebener_2016} \\
	C$_2$H$_4$Br$_2$ & 106-93-4 & Ethylene dibromide & \cite{Stoll_2003} \\
	C$_2$H$_4$Br$_3$ & 557-91-5 & 1,1-Dibromoethane & \cite{Stoll_2003} \\
	C$_2$H$_4$Cl$_2$ & 75-34-3 & 1,1-Dichloroethane & \cite{Stoll_2003} \\
	C$_2$H$_4$F$_2$ (R152a) & 75-37-6 & 1,1-difluoro-Ethane & \cite{Stoll_2003} \\
	C$_2$H$_4$O & 75-21-8 & Ethylene oxide & \cite{Eckl_2008a,Huang_2012} \\
	C$_2$H$_5$Br & 74-96-4 & Ethyl bromide & \cite{Stoll_2003} \\
	C$_2$H$_5$F & 353-36-6 & Fluoroethane & \cite{Stoll_2003} \\
	C$_2$H$_6$ & 74-84-0 & Ethane & \cite{Vrabec_2001,Vrabec_2006,Stoebener_2016} \\
	C$_2$H$_6$O & 115-10-6 & Dimethyl ether & \cite{Eckl_2008b} \\
	C$_2$H$_6$O & 64-17-5 & Ethanol & \cite{Schnabel_2005} \\
	C$_2$H$_6$O$_2$ & 107-21-1 & Ethylene glycol & \cite{Huang_2012} \\
	C$_2$H$_6$S & 75-18-3 & Dimethyl sulfide & \cite{Eckl_2008b} \\
	C$_2$H$_7$N & 124-40-3 & Dimethylamin & \cite{Schnabel_2008} \\
	C$_2$H$_8$N$_2$ & 57-14-7 & 1,1-Dimethylhydrazine & \cite{Elts_2012} \\
	C$_2$HBrClF$_3$ & 151-67-7 & Halothane & \cite{Stoll_2003} \\
	C$_2$HCl$_2$F$_3$ (R123) & 306-83-2 & 2,2-dichloro-1,1,1-trifluoro-Ethane & \cite{Stoll_2003} \\
	C$_2$HCl$_3$ & 79-01-6 & Trichloroethylene & \cite{Stoll_2003} \\
	C$_2$HClF$_2$ (R1122) & 359-10-4 & 2-Chloro-1,1-difluoroethylene & \cite{Stoll_2003} \\
	C$_2$HClF$_4$ (R124) & 2837-89-0 & 2-chloro-1,1,1,2-tetrafluoro-Ethane & \cite{Stoll_2003} \\
	C$_2$HF$_5$ (R125) & 354-33-6 & pentafluoro-Ethane & \cite{Stoll_2003} \\
	C$_2$N$_2$ & 460-19-5 & Cyanogen & \cite{Miroshnichenko_2013} \\
	C$_3$H$_4$ & 463-49-0 & Propadiene & \cite{Vrabec_2001} \\
	C$_3$H$_4$ & 74-99-7 & Propyne & \cite{Vrabec_2001} \\
	C$_3$H$_6$ & 115-07-1 & Propylene & \cite{Vrabec_2001} \\
	C$_3$H$_6$ & 75-19-4 & Cyclopropane & \cite{Munoz_2015} \\
	C$_3$H$_6$O & 67-64-1 & Acetone & \cite{Windmann_2014} \\
	C$_3$H$_8$O & 67-63-0 & Propan-2-ol & \cite{Munoz_2018} \\
	C$_3$HF$_7$ (R227ea) & 431-89-0 & 1,1,1,2,3,3,3-heptafluoro-Propane & \cite{Eckl_2007} \\
	C$_4$F$_{10}$ & 355-25-9 & Perfluorobutane & \cite{Koester_2012} \\
	C$_4$H$_{10}$ & 75-28-5 & Isobutane & \cite{Eckl_2008b} \\
	C$_4$H$_4$S & 110-02-1 & Thiophene & \cite{Eckl_2008b} \\
	C$_4$H$_8$ & 287-23-0 & Cyclobutane & \cite{Munoz_2015} \\
	C$_5$H$_{10}$ & 287-92-3 & Cyclopentane & \cite{Munoz_2015} \\
	C$_6$H$_4$Cl$_2$ & 95-50-1 & ortho-Dichlorobenzene & \cite{Huang_2011} \\
	C$_6$H$_5$Cl & 108-90-7 & Chlorobenzene & \cite{Huang_2011} \\
	C$_6$H$_6$ & 71-43-2 & Benzene & \cite{Huang_2011,Guevara-Carrion_2016} \\
	C$_6$H$_6$O & 108-95-2 & Phenol & \cite{Munoz_2017} \\
	C$_6$H$_7$N & 62-53-3 & Aniline & \cite{Munoz_2017} \\
	C$_6$H$_{10}$O & 108-94-1 & Cyclohexanone & \cite{Merker_2012b} \\
	C$_6$H$_{12}$ & 110-82-7 & Cyclohexane & \cite{Eckl_2008b,Merker_2012b,Munoz_2015} \\
	C$_6$H$_{12}$O & 108-93-0 & Cyclohexanol & \cite{Merker_2009} \\
	C$_6$H$_{13}$N & 108-91-8 & Cyclohexylamine & \cite{Munoz_2017} \\
	C$_6$H$_{18}$OSi$_2$ & 107-46-0 & Hexamethyldisiloxane & \cite{Thol_2016A} \\
	C$_7$H$_8$ & 108-88-3 & Toluene & \cite{Huang_2011,Guevara-Carrion_2016} \\
	C$_8$H$_{24}$O$_4$Si$_4$ & 556-67-2 & Octamethylcyclotetrasiloxane & \cite{Thol_2016B} \\ 
\end{longtable}

\section{Available Input Formats}

Questions regarding the molecular topology and the interaction specifications in a certain simulation program are among the most common ones in the mailing lists, for example for LAMMPS \cite{Plimpton_95} and Gromacs \cite{VanDerSpoel_2005}. The 'by hand' implementation of force fields is a sensitive task \cite{Schappals_2017} due to questions like angle definitions, units, and geometry specifications, etc. Ready-to-use and validated force field input files are -- where available -- a welcome support to simulators. The MolMod database currently provides input files for the simulation programs LAMMPS \cite{Plimpton_95}, Gromacs \cite{VanDerSpoel_2005}, \textit{ms}2 \cite{Rutkai_2017}, and \textit{ls}1 mardyn \cite{Niethammer_2014}. Since these molecular simulation codes are based on different approaches to realise for example multipole interactions, the geometry definition or the rigidity constraint of molecules, on the fly conversion routines were implemented in the database. 

Since most simulation programs cannot handle point dipoles or point quadrupoles explicitly, they are converted on the fly into the corresponding configuration of point charges, i.e ($+q$, $-q$) for dipoles and ($+q$, $-2q$, $+q$) for quadrupoles. Despite the fact that point multipoles are computationally much cheaper than a corresponding configuration of point charges, both are physically equivalent \cite{Engin_2011}. As there is one degree of freedom for replacing point dipoles and point quadrupoles by point charges, a rational choice for one parameter -- either the elongation or the magnitude of the point charge -- has to be made. This problem was addressed in the work of \textit{Engin et al.} \cite{Engin_2011} for the molecular model class of 2CLJD and 2CLJQ. \textit{Engin et al.} \cite{Engin_2011} could show that setting the distance $a$ or $d$ between the two point charges to $\frac{\sigma}{20}$, where $\sigma$ is the size parameter of the Lennard-Jones site, leads to excellent agreement between a multipole and the corresponding configuration of point charges. The magnitude of the two point charges $q$ for a dipole is then straightforwardly computed as
\begin{equation}
q=\frac{\mu}{a}=\frac{20\,\mu}{\sigma},
\end{equation}
for dipoles and 
\begin{equation}
q=\frac{Q}{2\,d^2}=\frac{200\,Q}{\sigma},
\end{equation}
for quadrupoles. This method was extended in the MolMod database to arbitrary molecular structures by calculating the parameter $a$ or $d$ by means of the Lennard-Jones interaction site closest to the point dipole according to the Euclidean norm. Also, the geometry of the molecular models is adapted to the demands of each simulation program, as some require internal Z-matrix coordinates and others \textit{xyz}-coordinates. Particular care was also taken regarding the units of each numeric value. 

Each molecular force field consists of a single and unique set of parameters that is converted on demand into the desired input file format. All data is stored in the database using SQL (structured query language) to ensure flexibility and robustness with the data handling. All types of molecular force fields in the MolMod database were validated for the supported simulation programs.

\section{Nomenclature}

A comprehensive description of the nomenclature for all supported input files is given to facilitate the integration of the force field input files into the workflow of a simulator. The input files for the different simulation programs that are provided by the MolMod database contain basically only numbers in the prescribed formats, sometimes in combination with keywords. As such they are ready-to-use. This is sufficient for standard applications of the corresponding simulation programs. But numbers that carry no semantics are insufficient for anything that goes beyond plug and play. Therefore, the MolMod database contains a comprehensive description of the content of the input files for all supported simulation programs and the force field definition in the MolMod web frontend. This description of meaning of the provided data enables using the data also in non-standard applications and ensures that the contents become genuinely interoperable with data and metadata from other sources, for example for the Virtual Materials Marketplace (VIMMP) where data pertaining to a variety of modelling paradigms need to be processed and made accessible to end users in a coherent way \cite{VIMMP_website}.

Appropriate semantic assets were defined for each entity of the database and eventually for their relations. The database was then complemented by an according nomenclature, i.e. a conceptual model of the employed entities in a descriptive form that is to be processed by humans, not by machines. This nomenclature, which is available on the MolMod web frontend, builds on - and is compatible with - the Review of Materials Modelling (RoMM) by \textit{de Baas} \cite{deBaas_2018_book}. It defines the sense in which the terms associated with the database entries are meant to be used, including, but not limited to:
\begin{itemize}
	\item geometry definitions (\textit{xyz}-coordinates, Z-matrix, angle and dihedral definitions, etc.)
	\item definition of multipole moments and equivalent point charge configurations 
	\item employed unit systems. 
\end{itemize}

\section{Summary and Future Work}

In the present work the MolMod database is introduced. It is a open accessible new database of molecular models of fluids and contains presently models of about 150 pure substances. The models were parametrised in previous work such as to yield a good description of experimental data of the vapour-liquid equilibrium of the studied fluid. Many of the models have also been tested with respect to caloric, transport and interfacial properties and the predictions were often found to be in good agreement with experimental data. This makes the models provided in the MolMod database interesting for applications in chemical engineering and many other fields of science and technology. 

The MolMod database provides ready-to-use input files for different common molecular simulation packages. This reduces the risk of input errors in the simulations. But the MolMod database also contains additional information: References to the original work in which the models were developed are given as well as known misprints clearly pointed out. Furthermore, the MolMod database also provides detailed descriptions of all entities in the input files that are available from the database. 

The MolMod database was developed and is maintained by the Boltzmann Zuse Society of Computational Molecular Engineering (BZS). The models that are presently included were developed and tested by BZS members, but the MolMod database is open for contributions from other groups. Any tailor-made molecular model of a fluid that fits into the scope of the MolMod database may be included provided that its development and testing have been described in a peer reviewed scientific paper and that a sufficiently accurate description of relevant thermophysical data is achieved. Suggestions to include new models can be filed via the MolMod homepage. 

In its present version, the MolMod database only contains models of simple fluids without internal degrees of freedom. An extension to large molecules that require a modelling with internal degrees of freedom is interesting. The same holds for an extension to the Mie potential instead of the Lennard-Jones 12-6 potential. Future work will also aim at providing input files for other molecular simulation programs than Gromacs, LAMMPS, \textit{ms}2, \textit{ls}1 mardyn, that are already served.

Ongoing and future developments will also advance the semantic technology that underlies to the MolMod database to include (or attach to) machine-readable assets, e.g. by rigorously relating the present nomenclature, that categorises and describes atomistic and mesoscopic materials models, to the MODA Workow Graph Language \cite{Schmitz_2018_agreement}, ontologies such as the European Materials Modelling Ontology \cite{Ghedini_2019}, and the metadata model EngMeta \cite{Schembera_2019}. This will facilitate integrating the MolMod database as a component into future development work towards open translation environments, i.e. platforms that generate materials modelling workflows that are tailored to the needs of the end user, and virtual marketplace platforms where materials modelling services and data access can be traded. Various approaches for providing such environments have so far been envisioned; cf. \textit{de Baas} \cite{deBaas_2018_book} for a detailed discussion of the position that semantic assets for molecular modelling and simulation, model repositories, and repositories for simulation data and metadata may occupy within a coherent approach to materials modelling interoperability and standardisation.


\section*{Acknowledgement}

We gratefully appreciate running test simulations for the validation of the GROMACS, \textit{ls1} mardyn and LAMMPS input files by Matthias Heinen, Edder J. Garcia and Angelo Damone, respectively. The simulations for the parametrisation of the force fields were carried out on the HazelHen at High Performance Computing Center Stuttgart (HLRS) under the grant MMHBF2 as well as on the SuperMuc at Leibniz Supercomputing Centre (LRZ) Garching under the grant SPARLAMPE (pr48te). The present research was conducted under the auspices of the Boltzmann-Zuse Society of Computational Molecular Engineering (BZS).

\section*{Funding}

The authors gratefully acknowledge funding of the present work by the Reinhart Koselleck program of the Deutsche Forschungsgemeinschaft (DFG), and the TaLPas project by the German Federal Ministry of Education and Research (BMBF). They acknowledge funding by the "Horizon 2020" research and innovation programme of the European Union (EU): The co-authors S.~Stephan and H.~Hasse by grant agreement No. 694807 -- ENRICO, and the co-author Martin Horsch by grant agreement No. 760907 -- Virtual Materials Marketplace (VIMMP).

\FloatBarrier
\newpage

\FloatBarrier

\end{document}